# 0.001% and Counting: Revisiting the Price Rounding Tax*

Doron Sayag
Department of Economics, Bar-Ilan University
Ramat-Gan 5290002, Israel,
Israel Central Bureau of Statistics
dorons@cbs.gov.il

Avichai Snir
Department of Economics, Bar-Ilan University
Ramat-Gan 5290002, Israel
Avichai.Snir@biu.ac.il

Daniel Levy
Department of Economics, Bar-Ilan University
Ramat-Gan 5290002, Israel,
Department of Economics, Emory University
Atlanta, GA 30322, USA,
ICEA, ISET at TSU, and RCEA
Daniel.Levy@biu.ac.il

Revised:

November 5, 2025

* We are grateful to the anonymous reviewer for offering constructive comments and suggestions, and to the editor Craig Depken for guidance. We thank the participants of the 6[th] Annual Conference on Law and Macroeconomics, held on November 2–3, 2024, at Tulane Law School, Tulane University, for their helpful and constructive comments and suggestions. The paper is based on Chapter 2 of Doron Sayag's PhD dissertation at Bar-Ilan University. All errors are ours.

**Correspondence**: Avichai Snir. Email: Avichai.Snir@biu.ac.il

**Abbreviations**: NIS, New Israeli Shekel; CBS, Central Bureau of Statistics; FMCG, Fast Moving Consumer Goods; UPC, Universal Product Code; CAD, Canadian Dollar.

# 0.001% and Counting:
# Revisiting the Price Rounding Tax

### *Abstract*

In 1991 and 2008, Israel abolished the equivalents of 1¢ and 5¢ coins, respectively, effectively eliminating low-denomination coins and introducing rounding in cash transactions. When totals were rounded up, shoppers incurred a small *rounding tax*. Using detailed data on price endings and basket sizes across supermarkets, drugstores, small groceries, and convenience stores, we estimate that the magnitude of the rounding tax borne by Israeli consumers averaged only 0.001%–0.002% of revenues in the fast-moving consumer goods markets. These findings have implications for the ongoing debate regarding the desirability and viability of abolishing the 1¢ and 5¢ coins in the US.

"Most pennies produced by the U.S. Mint are given out as change but never spent; this creates an incessant demand for new pennies to replace them, so that cash transactions that necessitate pennies (i.e., any concluding with a sum whose final digit is 1, 2, 3, 4, 6, 7, 8, or 9) can be settled. Because these replacement pennies will themselves not be spent, they will need to be replaced with new pennies that will also not be spent, and so will have to be replaced with new pennies that will not be spent, which will have to be replaced by new pennies (that will not be spent, and so will have to be replaced). In other words, we keep minting pennies because no one uses the pennies we mint."

**Caity Weaver, "Stop Making Cents,"**
*New York Times Sunday Magazine*, **September 8, 2024, p. 26**

"On January 3, 2012, [Starbucks] set the net price of a tall cup of coffee in Manhattan at precisely $2.01, including tax. "I didn't have the penny." … Ms. Schmais wasn't especially irked by the price increase, which comes to 10 cents for a tall cup. But that orphaned penny had her fuming. "It's the stupidity of it," she said, "It's what I'd call 'the annoyance factor.'…It's ridiculous. Why the extra penny? Who has pennies? Didn't anyone think this through? Couldn't they round down or even up? Why leave it at a penny?" … When David Turnbull presented two $1 bills for coffee at the Starbucks on Astor Place, he met the same problem. He wasn't carrying any pennies. "I can't believe it," he said. "Now I need to walk around with pennies? Who could possibly think a price of $2.01 makes sense?"

**Jeff Sommer, "Dear Starbucks: A Penny for Your Thoughts,"**
*New York Times*, **January 15, 2012, p. BU3**

## 1 | INTRODUCTION

When small denomination coins, such as the 1¢, are eliminated, the currency becomes less divisible, leading to situations where prices, such as $9.99, cannot be paid exactly. In such situations, the total amount due is rounded up or down in cash transactions to the nearest amount that can be paid precisely. When bills are rounded up, shoppers might end up paying a few pennies more than they should. This process leads to concerns that the elimination of small-denomination coins might result in shoppers paying a significant amount of *rounding tax*. Indeed, according to "Americans for Common Cents," 73% of the public worries about the rounding tax.[1]

Our goal is to assess quantitatively the size of the rounding tax. In 1991 and 2008, Israel abolished the NIS 0.01 and NIS 0.05 coins (NIS is the New Israeli Shekel, the currency of Israel), respectively, because of their high cost of production.[2] Although the NIS 0.01 and NIS 0.05 coins have not been in use since then, until 2014, retailers were free to use any price ending when setting their prices. In credit card transactions, the amount paid was exact.

In cash transactions, however, the total bill was rounded to the nearest NIS 0.05 until January

---

[1] Source: Americans for Common Cents, an organization lobbying against abolishing the pennies in the US, https://pennies.org/ (accessed September 18, 2025).

[2] For example, the cost of minting a NIS 0.05 coin was NIS 0.16. Also, the public was reluctant to accept them as a change. Coin-operated devices stopped accepting them as well. See https://www.boi.org.il/en/information-and-service-to-the-public/my-cash-banknotes/the-bank-of-israel-asks-the-government-to-approve-abolishing-the-5-agorot-coin-as-legal-tender/ (accessed September 18, 2025).

2008, and to the nearest NIS 0.10 between 2008 and 2014.[3] The rounding rule used between 2008 and 2014 drew consumer criticism for its asymmetry: 4 endings (1–4) were rounded down, while 5 endings (5–9) were rounded up (Levy et al. 2011).

In October 2013, the government announced a new price rounding regulation: starting January 1, 2014, *all prices* must end in 0, and thus, thereafter, there would no longer be a need to round individual transaction bills.[4] We assess the size of the rounding tax that Israeli shoppers were paying right before the new regulation came into effect, using the methodology of Lombra (2001), Chande and Fisher (2003), and Whaples (2007).

The question of rounding tax is important for policymakers because many countries are considering stopping the minting of low-denomination coins for various reasons. This is particularly true in the US, where the feasibility and desirability of eliminating the 1¢ coin has been debated for a long time. Having an accurate estimate of the rounding tax that the public may end up paying if the 1¢ coin is indeed eliminated is important for making an informed decision on this matter.

A key advantage of our paper over earlier work is the richness of our dataset. First, our dataset includes information on the *actual* distributions of both the price endings and the shopping basket sizes, which were not available to the above authors. Second, we are able to measure the rounding tax separately for different store types. As a result, our rounding tax estimates are more precise. In addition, our rounding tax estimates capture the heterogeneity that is found across different types of retail stores and formats (Ray et al, 2023).

We find that the rounding tax paid by shoppers in Israel was quite small, in the range of about 0.001%–0.002% of the total revenues in the fast-moving consumer goods market. This finding is important, as it suggests that the rounding tax shoppers pay for the absence of low-denomination coins is substantially smaller than earlier studies have reported. Perhaps more importantly, the finding has implications for the ongoing public debate in the US concerning the desirability and viability of abolishing small denomination, the 1¢ and 5¢, coins.

---

[3] Similar ("Swedish") rounding rules were adopted in other countries that abolished small denomination coins, including Sweden, New Zealand, Australia, Finland, Canada, and the Netherlands (Leszkó 2009). The posted prices in Israel, quoted in NIS, are final because they include all taxes.

[4] Mr. N. Benett, who was the Minister of the Economy (and later became the Prime Minister of Israel), defended the regulation: "For years, retailers profited twice—they both [1] presented a lower price that misled the consumers and, [2] collected an excess amount from consumers." Source: https://www.kolhaemet.co.il/4174 (in Hebrew), accessed on September 18, 2025. Mr. Benett's point [1] refers to 9-ending prices (Schindler and Kibarian 1996, Anderson and Simester 2003, Thomas and Morwitz 2005, and Snir and Levy 2021), while his point [2] refers to the asymmetric price rounding rule that was in use during that time period. We shall note that the price rounding law doesn't apply to goods sold by weight and to 21 basic food products with capped prices (Sayag et al. 2025).

The paper is organized as follows. In the next section, we review the literature. In section 3, we describe the data. In section 4, we estimate the rounding tax. In section 5, we assess the generalizability and lessons of our findings, followed by conclusions and caveats in section 6.

## 2 | LITERATURE REVIEW

The limited literature that exists on rounding tax is primarily motivated by the debate on the merits of eliminating the 1¢ coin in the US. Lombra (2001) argues that shoppers will incur a financial penalty because the final bills would be rounded up in the absence of low-denomination coins. Using convenience store data from the US, he estimates the rounding tax that the US shoppers would pay if the 1¢ coin were abolished, and consequently, cash transactions were rounded to the nearest 5¢. Lombra assumes that (i) shoppers typically buy 1–2 items, (ii) 82.5% of the prices end in 9, and (iii) 50%–83% of the transactions are in cash. Based on these assumptions, his estimate of the annual rounding tax is approximately $700 million.

Keinsley (2013) finds that eliminating the 1¢ coin would cost US shoppers up to $1.6 annually on average because of rounding, yielding a total rounding tax of roughly $530 million annually. Whaples (2007) uses data from a convenience store chain in 7 US states. After accounting for local sales taxes, he finds that in 6 of the 7 states, the rounding tax would actually benefit the US consumers. The predicted gains, however, are small, in the range of about 0.01¢–0.10¢ per cash transaction.

Evidence from other countries is broadly in line with the above findings. For example, Cheung (2018) finds that since Canada abolished the 1¢ coin in 2013, the Canadian shoppers lost CAD 3.3 million per year. This, Cheung (2018) shows, amounts to CAD 157 extra revenue for an average Canadian grocery store annually. Similarly, Chande and Fisher (2003) study price data from a Canadian fast-food chain. They assume that shoppers buy 1–4 items on average, and after adding sales tax to the posted prices, they find that the rounding tax is positive but small and inconsequential. Finally, Leszkó (2009) surveys in great detail the countries that abolished low-denomination coins. She also concludes that their effect on the rounding tax is small.

## 3 | DATA

We use three datasets. First, we use the Israeli Central Bureau of Statistics (*CBS*) CPI micro-level dataset for the 2005–2014 period. The dataset includes product-level prices collected at a representative sample of outlets that belong to convenience stores, drugstores, supermarkets, and small grocery stores. We focus on Fast-Moving Consumer Goods (*FMCG*)—products that are frequently bought, quickly consumed, have low prices, and are sold in large quantities, because these products are most likely to be relevant for rounding tax. We limit the analysis to the products with a maximum price of NIS 200 (equivalent to about US$60).[5] This yields a total of 4,341,643 monthly observations. The average and median prices in the sample are NIS 28.85 and NIS 14.29, respectively.

Second, we utilize the Nielsen dataset for the period 2011–2014, which contains monthly observations on revenues and sales volume of products in FMCG categories. The data, which covers 96% of the FMCG market, is at the national level and is divided into three categories based on retailer type: supermarkets and drugstores, convenience stores, and small groceries. The data includes information on the total sales volume and revenues for a product by store type.[6]

Third, we incorporate data on Israeli consumers' shopping habits from the 2013 CBS Household Expenditure Survey. The data set comes from bi-weekly diaries kept by the representative 9,500 households, listing every product they have purchased over the relevant 14-day period, including the store type and the purchase price. In total, the dataset comprises approximately 870,000 item-level observations. We use this data to determine the number of items purchased per shopping trip.

## 4 | ESTIMATES OF THE ROUNDING TAX

### 4.1 | Distribution of price endings

The CPI data contains information on the prices of all the products sampled by CPI surveyors in 2013. We use it to calculate the shares of price endings by type of store: small grocery stores, supermarkets, drugstores, and convenience stores.[7] We compute the shares for 2013 because this is the final year before the adoption of the price rounding regulation. Thereafter, the rounding tax is irrelevant because the regulation banned non-round prices altogether.

The left-hand-side panel of Figure 1 shows the distributions of the price endings. There is large heterogeneity in the distributions of both the price endings and shopping baskets across store types. Consistent with Knotek (2011), Levy et al. (2020), and Snir et al. (2022), 9 is the most common price ending in supermarkets and drugstores, while 0 is the most common price ending in small grocery stores and convenience stores. In supermarkets and drugstores, 9-ending and 0-ending prices comprise 61.1% and 18.3%, in small grocery stores 19.1% and 75.8%, and in convenience stores 34.2% and 64.1% of the prices, respectively.

---

[5] Levy et al. (2011) show that as prices get higher, the 9-digits tend to shift to the left, implying that the right-most digits of relatively high prices are usually zero.

[6] For illustration, an observation may state that on a given month, a particular brand of orange juice, identified by its Universal Product Code (UPC), sold in convenience stores 19,543 units, for a total revenue of NIS 140,123.

[7] We do not separate the information on supermarkets and drugstores because Nielsen treats them as a single group.

## 4.2 | Distribution of the basket size

Using information from the bi-weekly diaries of the 2013 Household Expenditure Survey, we calculate the number of items bought on each shopping trip, and identify the type of store— supermarket, small grocery, convenience store, or drugstore— where the items were purchased. We then construct the distribution of basket sizes by store type.

The right-hand-side panel of Figure 1 shows the shopping basket sizes by store types. In convenience stores, 71.0% of baskets contain a single item, similar to the figures reported for the US (Lombra 2001) and Canada (Chande and Fisher 2003). In supermarkets and drugstores, 13.6% of the baskets contain one item, the median basket contains 6 items, and 25% of the baskets contain 15 or more items. Small grocery stores, with a median basket size of 3 items, fall in between.

## 4.3 | Simulation

To estimate the rounding tax, we use the distributions of price endings and basket sizes to simulate 10,000 transactions for each type of retailer (Lombra 2001, Chande and Fisher 2003). In each simulation, we employ a two-stage procedure. In the first stage, we determine the size of the shopping basket based on the likelihood of observing each basket size. For example, for convenience stores, we determine the size of each of the 10,000 transactions by randomly drawing a basket size according to the probabilities given in Figure 1. For example, a basket with 1 item occurs with a probability of 71.0%, a basket with 2 items occurs with a probability of 16.5%, and so on.

In the second stage, for each product in the basket, we randomly assign a price ending based on the distribution for that store type. For instance, if a simulated convenience store basket contains 2 items, the price ending of the first product is drawn from a distribution where a 0-ending occurs with 64.1% probability, a 1-ending with 0.04%, etc., as reported in Figure 1. After assigning price endings to all products in the basket, we calculate the rounding tax for the transaction. For example, a basket with two 9-ending items generates a rounding tax of NIS 0.02. Finally, we compute the average rounding tax across all 10,000 simulated transactions.

## 4.4 | Results

Column 1 of Table 1 reports the simulation results for the average rounding tax per cash transaction for each store type. The average rounding tax is the highest in drugstores and supermarkets, with NIS 0.0075 per transaction, and the lowest in convenience stores, with NIS 0.0048 per transaction. In small grocery stores, the tax averages NIS 0.0058 per transaction.

Column 2 presents the share of each type of store in the total FMCG revenue in 2013, based on Nielsen data. Column 3 reports the total number of FMCG transactions in 2013 by type of store, calculated by dividing the total number of transactions for each store type (from the Nielsen data) by the average number of items per shopping trip in the corresponding type of store.

If shoppers paid a rounding tax for all transactions, we could calculate its total amount by store type by multiplying columns (1) and (2). However, the rounding tax applies only to cash transactions. To control for this, we use data on total expenditure and payments by credit card in the FMCG market. We estimate that the share of cash transactions in FMCG transactions in 2013 was 25%.[8]

Since we do not have data on credit card use by store type, Column 4 reports our baseline estimate of the total rounding tax for 2013, assuming that the cash transactions share was 25% in all stores. This yields a total rounding tax estimate of NIS 507,280 in 2013.

However, it is unlikely that the share of cash transactions in supermarkets was the same as in convenience stores.[9] We therefore estimate the upper and lower bounds by varying the share of cash transactions across store types, so that the overall share of cash transactions remains 25%. We then find the shares of cash transactions that maximize and minimize the overall rounding tax.

We find that the maximum (minimum) rounding tax is obtained when the share of cash transactions in small grocery stores and convenience stores is 100% (0%), implying that the share of cash transactions in supermarkets and drugstores is 10.6% (29.8%).[10] The total rounding tax paid in 2013 in the maximum (minimum) scenario is NIS 763,641 (NIS 422,390). According to Nielsen's data, the total revenue in the FMCG market in 2013 was NIS 40.8 billion, implying that the rounding tax in 2013 was in the range of 0.001%–0.002% of the total revenue.

---

[8] We estimate the share of cash transactions in the FMCG market by taking the difference between total expenditures (from Nielsen data) and total credit card transactions (from the CBS data).
[9] Bouhdaoui et al. (2014), Chen et al. (2019), and Shy (2020) show that the share of cash transactions is higher in convenience stores than in supermarkets and drugstores.

## 5 | DISCUSSION: GENERALIZABILITY AND LESSONS FROM OUR FINDINGS

There are interesting questions that concern the generalizability of the findings we are reporting in this paper, and the lessons they offer. To address these questions, consider the settings and the considerations that lead countries to eliminate low-denominational coins.

### 5.1 | Reasons for the elimination of low-denomination coins

Several reasons prompted the Bank of Israel to eliminate the low-denomination coins. First, the coins had little or no purchasing power. Second, the cost of producing the coins exceeded their nominal face value. Third, beyond the direct cost of production, there were additional costs, including the costs of processing, storing, handling, circulating, transporting, etc., which further increased their economic cost. Fourth, vending machines, parking meters, and other coin-operated devices were no longer accepting them. Fifth, both shoppers and cashiers were inconvenienced by the need to handle them.

These considerations are not unique to Israel. Indeed, according to Leszkó (2009), the countries that have decided to eliminate low-denomination coins did so for similar reasons. The reports produced by central banks and/or minting authorities of countries that have either abolished or consider abolishing low-denomination coins emphasize factors closely aligned with those that led the Bank of Israel to withdraw the NIS 0.01 and NIS 0.05 coins.[11] The reasons cited in the US in support of eliminating the 1¢ coin are quite similar. First, the cost of producing and distributing a 1¢ coin far exceeds the coin's nominal face value. As Figure 2 shows, the gap between the nominal face value of the coin ("1¢") and the actual cost of producing and distributing the coin has increased (almost monotonically) from 0.97¢ in 2005 to 3.69¢ in 2024, which is reflected in the corresponding *negative* seigniorage figures (the $-value of the gap between the face value of a coin, and its cost of production and distribution). For example, in 2021, the seigniorage for the 1¢ coin was –$83.6 million, and in 2024 it was –$85.3 million.[12]

---

[10] Note that when we calculate the maximum rounding tax, we look for the maximum total rounding tax. Therefore, when we calculate the maximum rounding tax, we have a high (low) share of transactions in small groceries and convenience stores (supermarkets and drugstores). Thus, the maximum rounding tax is obtained when the rounding tax in supermarkets and drugstores is lower than in the baseline scenario, while the rounding tax in grocery stores and convenience stores is higher than in the baseline (equal shares) scenario. We obtain the opposite pattern when we calculate the minimum rounding tax.

[11] See, for example, the report of the Royal Australian Mint, at https://www.ramint.gov.au/collect/national-coin-collection/circulating-coins/one-cent (accessed September 18, 2025), which explains the reasons for eliminating 1¢ and 2¢ coins: "The cessation of issue of one and two cent coins was announced by the Treasurer in his Budget Speech of 21 August 1990. The decision was based on the loss of real purchasing power through inflation and the cost of minting these coins." As another example, see the report of the Bank of Canada, at https://www.bankofcanadamuseum.ca/2025/08/whatever-happened-to-the-penny-a-history-of-our-one-cent-coin/

Second, it is widely recognized that Americans hoard pennies, necessitating minting them again, and again, year after year, as expressed in the opening quotation from Caity Weaver's September 10, 2024 *NYT Sunday Magazine*. Third, the low purchasing power of the 1¢ coin is often cited in debates about the desirability of eliminating the coin. Fourth, the inconvenience caused by the need to handle 1¢ coins in cash transactions annoys many American shoppers. For example, the second quotation, also from the *NYT*, that starts this paper, offers a good example of the strong negative feelings and emotions that dealing with pennies elicits from the US shoppers. Fifth, the increasing use of other means of payment, such as credit cards, digital wallets, etc., instead of cash in the US, has been cited as another reason in support of eliminating the 1¢ coin.

## 5.2 | Implementation in Israel

While the main reasons for the elimination of the low-denomination coins in Israel were similar to the reasons listed above (their high cost of production, inability to use them with vending machines, parking meters, and other coin-operated devices etc.), its implementation in Israel was done differently, in comparison to other countries. The rounding rule that was adopted by the Bank of Israel following the decision to eliminate the 5-agora coin from circulation was asymmetric, favoring the sellers. Out of 9 possible endings, prices with four endings (1, 2, 3, and 4) were rounded downwards, while prices with five endings (5, 6, 7, 8, and 9) were rounded

---

(accessed September 18, 2025), which explains the reasons for its 2012 decision to eliminate the Canadian penny: "Canada stopped issuing pennies for purely economic reasons. At the time of its demise, each cent cost 1.6¢ to make. This would be reason enough to cease production of any currency. But there was also little you could buy with one cent... By 2012, it was primarily used to make exact change." Similar explanations and reasonings are offered by the central bank and/or mint authorities of other countries.

[12] The sources of the seigniorage figures are *Coin News*, February 9, 2024, https://www.coinnews.net/2024/02/09/penny-costs-3-07-cents-to-make-in-2023-nickel-costs-11-54-cents-us-mint-realizes-249m-in-seigniorage/, and *Coin World*, January 24, 2025, https://www.coinworld.com/news/us-coins/u-s-mint-reveals-production-costs-for-2024, (both accessed September 18, 2025).

upwards. This seemed to benefit (even if marginally) the sellers, at the expense of the consumers.

We have not seen a similar asymmetric rounding mechanism employed in other countries. While we do not know the exact reasons for the Bank of Israel's decision to adopt an asymmetric price rounding rule that benefited the sellers, it might perhaps explain the unusual decision of the Israeli government to outlaw all non-zero ending prices. This decision was made under pressure from the Israeli public and consumer groups, who resented the asymmetric price rounding rule. While there are countries where most prices are round, as far as we know, it is not by decree, but rather by choice, primarily for the convenience that round prices offer in conducting market exchange.

### 5.3  |  Voluntary price rounding mechanisms

Some countries have introduced a voluntary price rounding mechanism. For example, the Irish Central Bank has not eliminated the 1¢ or 2¢ coins, which remain legal tenders. Instead, in 2015, the Bank established a price rounding rule to reduce the use of these coins in cash transactions because of their high cost of production relative to their nominal face value: the cost of producing a 1¢ coin was 1.65¢, while the cost of producing a 2¢ coin was 2.1¢. The symmetric rounding the Bank introduced is optional, and shoppers have the right to receive exact change upon request.[13]

Several other European countries have adopted similar voluntary price rounding mechanisms, while, in parallel, keeping the low-denomination coins as legal tender. If such rounding is widely adopted in these countries for most cash transactions, then to the extent that their settings and institutional features are similar to Israel, our estimates of the rounding tax are likely to be directly relevant for these countries as well.

### 5.4  |  Lessons and Generalizability

While Israel's asymmetric rounding rule makes it an unusual case, the lessons it offers are still relevant for other countries, particularly for the US, where the question of eliminating the penny is currently debated. First, the distribution of the last digit endings of retail prices in Israel, before the outlawing of the non-zero ending prices, was quite similar to the corresponding figures in the US. For example, Ater and Gerlitz (2017) find that in their dataset, which comes from Israeli supermarkets, 60% of the prices were 9-ending before such prices were outlawed in 2014. Snir et al. (2017) and Levy et al. (2020) use different datasets, also from Israeli supermarkets, and report that between 65% and 72% of the prices were 9-ending.

---

[13] Source: "Rounding of Cash Transactions," Central Bank of Ireland, https://www.centralbank.ie/consumer-hub/rounding (accessed September 18, 2025).

Corresponding figures reported for the US are in the same ballpark. For example, Levy, et al. (2011) report that in Dominick's, a large supermarket chain that was operating in the Chicago metro area (until it was acquired by Safeway), about 65% of the prices were 9-ending. Assuming that the shopping habits of the Israeli and US shoppers do not differ dramatically, this suggests that the *relative* magnitude of the rounding tax for US shoppers in the case of abolishing the 1¢ coin is likely to be similar to the estimates derived from the Israeli data.[14]

Second, according to our findings, in 2013, the total rounding tax Israeli shoppers ended up paying was in the range of 0.001%–0.002% of the total revenue. Given that the Israeli population in 2013 was 8.1 million, it implies an average annual rounding tax of NIS 0.05–NIS 0.09 per person, equivalent to about 1.4¢–2.5¢ (based on the yearly average exchange rate of $1 = NIS 3.6097). This is a significantly smaller figure than the $1.6 estimate by Keinsley (2013) for the US, if the 1¢ coin is indeed eliminated. The fact that we find such a small rounding tax under an asymmetric price rounding rule suggests that if countries adopt a symmetric price rounding rule, then the rounding tax is expected to be even lower, making our estimate effectively an upper bound.

An additional reason that our assessment of the rounding tax is likely to be an upper bound of the true rounding penalty is the increasing use of digital payment methods rather than cash. This is important because the rounding tax is primarily relevant for cash transactions, as rounding is applied in most cases only when the payment is made in cash. In this context, it is worth noting the recent trend in the US. The US Federal Reserve has been conducting an annual Diary of Consumer Payment Choice Survey since 2016 to track the shoppers' payment habits. According to the 2025 Report, the share of transactions made in cash in the US has dropped from 31% in 2016 to 14% in 2024, as Figure 3 shows. This suggests that the expected rounding tax has been declining over time, as a decreasing share of transactions involves cash.[15]

---

[14] We emphasize "relative" because the US economy is multiple times larger than the Israeli economy.

[15] Since 2016, the Federal Reserve Bank of Atlanta's Research Department has conducted the *Diary of Consumer Payment Choice* survey each year to gain insights into how people in the U.S. pay for goods and services. As noted by *Federal Reserve Financial Services* (2025), the survey collects detailed data from thousands of individuals who record every payment they make over three days in October—a month chosen to minimize the influence of seasonal spending. In the 2025 edition, 5,583 respondents participated, providing insights into their payment preferences,

On the other hand, household income is correlated with consumers' choice of payment instruments. According to the Federal Reserve Financial Services (2025, p. 13), credit availability, budgeting practices, access to liquidity, payment preferences, and access to banking services can limit the availability of certain payment instruments. Figure 4 shows that as household income increases, the shares of cash and debit card payments decrease, while the share of credit card payments increases. According to the figure, in 2024, American consumers earning less than $25,000 per year relied on cash for 24% of their payments, whereas those earning more than $150,000 used cash for only 9% of their payments.

This implies that lower-income households make more cash transactions than higher-income households, suggesting that it is likely that lower-income households will bear a large share of the rounding tax if the US penny is eliminated. Nevertheless, 24% of payments is similar to the average figure in Israel in 2013, 25%. Our findings suggest, therefore, that even among low-income households, the annual rounding tax is likely to be less than 2.5¢.

## 6 | CONCLUSION AND CAVEATS

In Israel, the NIS 0.01 and NIS 0.05 coins were abolished in 1991 and 2008, respectively. Until January 2014, the retailers were free to set prices with any ending, and in cash transactions, the final bills were rounded up or down to the nearest available currency denomination.

On January 1, 2014, the government enacted a "price rounding regulation" outlawing all non-0-ending prices. We calculate the rounding tax, defined as the additional amount shoppers paid due to price rounding upwards.

Our estimate of the rounding tax is 0.001%–0.002% of the total revenue in the FMCG market, which seems relatively small. The fact that, following the elimination of the NIS 0.01 and NIS 0.05 coins, an upward-biased asymmetric price rounding rule was adopted (5 price endings were rounded up, while 4 price endings were rounded down), suggests that the rounding tax would be even lower if only the 1¢ coin were abolished. In that case, rounding would have been naturally symmetric: with 4 price endings rounded up and 4 price endings rounded down.

---

including the use of cash, cards, checks, and digital methods. Participants also share information about the amount of cash they carry or store, and any deposits or withdrawals they make. The Dornsife Center for Economic and Social Research at the University of Southern California, which administers the survey, ensures that the data accurately reflects the behavior of the broader U.S. population by applying weights based on the statistics of the Census Bureau. Additional information is available in the survey's technical documentation and codebook accessible at the website of the Federal Reserve Bank of Atlanta, https://www.atlantafed.org/banking-and-payments/consumer-payments/survey-and-diary-of-consumer-payment-choice (accessed September 18, 2025).

In addition, when the price rounding regulation came into effect in January 2014, about 25% of all transactions in the FMCG market were paid in cash. The advances in electronic payment methods since 2014 have likely led to a decrease in the share of cash transactions, implying that the rounding tax is becoming smaller over time.

These findings and observations have important policy implications for the ongoing debate over the desirability and feasibility of abolishing the 1¢ coin in the US. According to the *New York Times* report on February 9, 2025, President Trump ordered the Treasury to stop minting pennies.[16] Our findings, along with the decline in cash usage for transactions, suggest that the rounding tax that American consumers will incur if the 1¢ coin is indeed abolished is likely to be small. Given the high production cost of the 1¢ coin, as Figure 2 shows, our findings suggest that the overall effect of eliminating the 1¢ coin would likely increase the overall social welfare.

We should note several caveats. First, while our price and shopping basket data come from sample surveys designed to be representative of the population, it is always possible that some population groups are over- or under-represented. We believe, however, that any such biases, if they exist, are likely to be small, because these surveys are routinely used for national statistics. Second, our Israeli data on credit card usage is limited. For example, we do not have information on credit card use by store type or by income level. Consequently, the results we report for the Israeli data cannot be used to assess which households bear the main burden of the rounding tax.

On the benefit side of keeping the 1¢ coin, some effects are hard or even impossible to measure and quantify. For example, our data cannot shed light on one of the benefits of keeping the low-denomination coins—currency divisibility, which diminishes the extent of price rigidity by making small price changes impossible.[17] Similarly, our data cannot speak to the emotional aspects of some people's resistance to eliminating low-denomination coins for the sentimental value they attach to the coins, as this effect seems impossible to quantify.

---

[16] Yan Zhuang and Erica Green, "Trump Orders Treasury to Halt the Minting of Pennies," *The New York Times* (New York Edition)*,* Section B, page 4, February 9, 2025.

[17] Levy and Young (2004) report that one of the reasons for the 70+ year-long nominal price rigidity of the Nickel (i.e., 5¢) Coke from 1886 to 1959 was the lack of low-denomination coins. They report that the Coca-Cola Company, along with other big soft drink manufacturers (including Pepsi-Cola, Quality Beverage, Royal Crown, Bireley's, etc.), wanted to increase the price of their drink while making it possible to continue using a single coin for purchasing the drinks. The companies, therefore, lobbied with the US President to have the US Treasury mint 7½¢ and 2½¢ coins, which would enable the Company to increase the price by 50% from 5¢ to 7½¢, while still using a single coin for purchasing it. The Company also considered asking for the minting of a 3¢ coin, which would enable it to hike the price from 5¢ to 6¢, while still using only one type of coin (two 3¢ coins). The US Treasury, according to Levy and Young's (2004) account, declined the Coca-Cola Company's request.

**DATA AVAILABILITY STATEMENT**

A Python code for implementing the simulation, together with a data file containing detailed information on the distributions of both the basket-size and price-endings, as well as the data used in plotting the figures, is available at:

Sayag, Doron, Snir, Avichai, and Levy, Daniel. (2025) Contemporary Economic Policy - Replication Package for "0.001% and Counting: Revisiting the Price Rounding Tax." Ann Arbor, MI: Inter-university Consortium for Political and Social Research [distributor], 2025-11-01. Available at: https://doi.org/10.3886/E239501V1.

The three datasets we use in the analysis (as described in section 3) were obtained from Israel's Central Bureau of Statistics (CBS). These data are subject to legal and administrative restrictions and must be requested directly from the CBS (https://www.cbs.gov.il). For more information, contact Doron Sayag at CBS, dorons@cbs.gov.il.

**FUNDING STATEMENT**

No funding was received for this project.

**CONFLICT OF INTEREST**

We have no conflict of interest.

**ORCID**

*Doron Sayag*  https://orcid.org/0000-0002-1116-9579

*Avichai Snir*  https://orcid.org/0000-0002-0957-2668

*Daniel Levy*  https://orcid.org/0000-0001-5740-3378

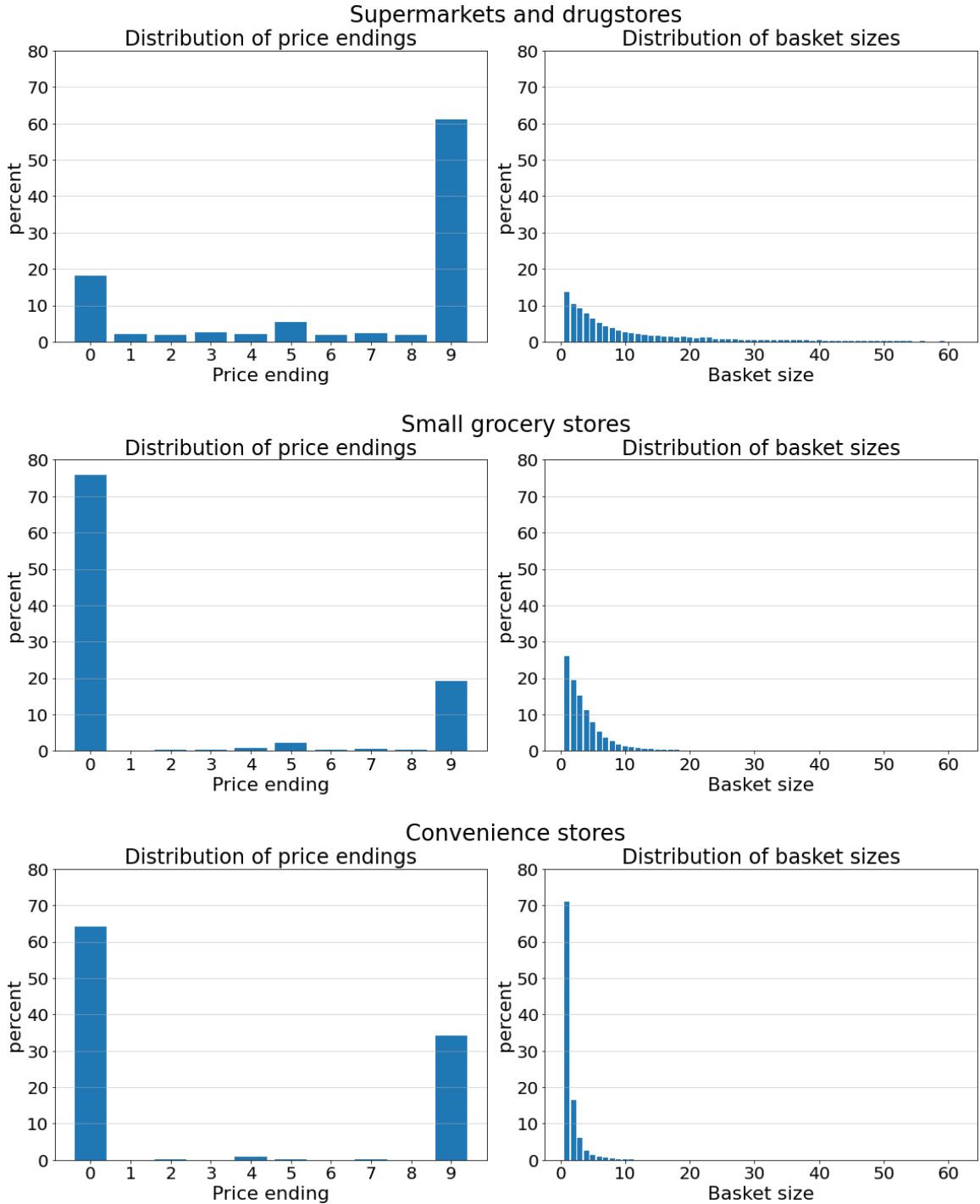

FIGURE 1   The distributions of price endings and basket sizes in 2013.

*Notes*: The left-hand-side panel depicts the distributions of the right-most digits in supermarkets and drugstores, small grocery stores, and convenience stores. The right-hand-side panel depicts the distribution of the shopping basket sizes in supermarkets and drugstores, small grocery stores, and convenience stores.

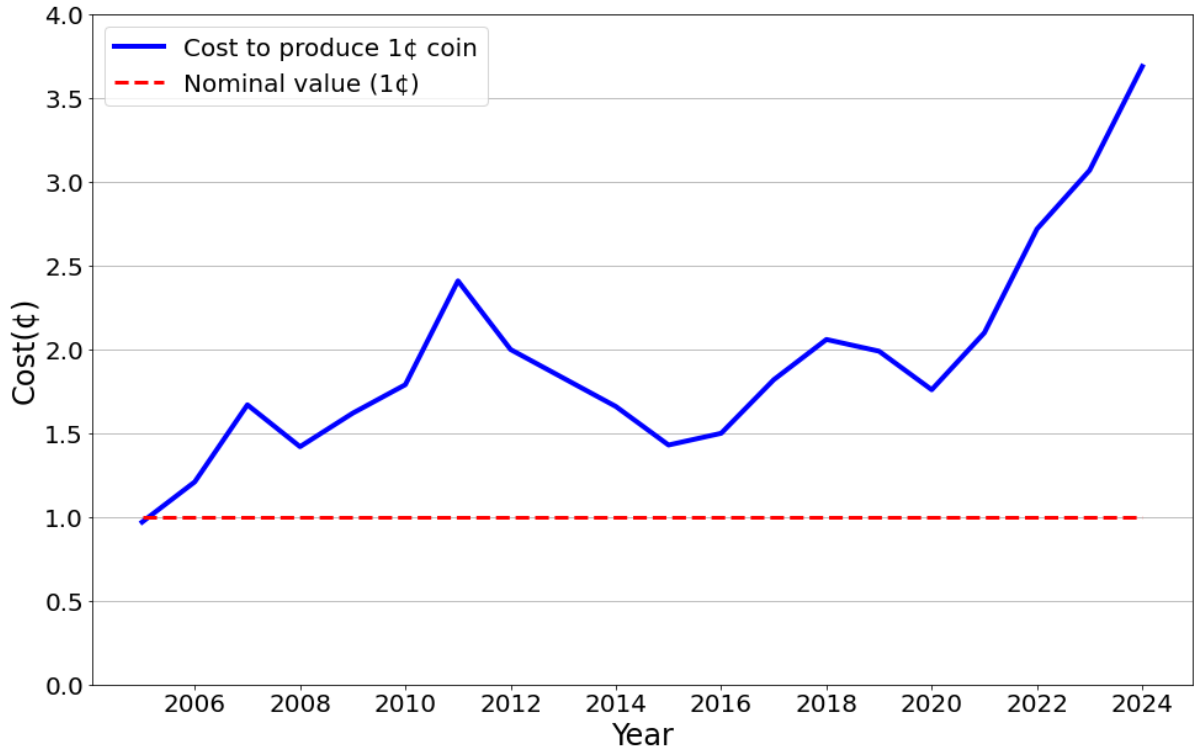

FIGURE 2   Cost of producing and distributing a 1¢ coin and its nominal face value.

*Note*: The source of the data presented in this figure is the Annual Reports of the United States Mint, https://www.usmint.gov/about/reports (accessed September 18, 2025).

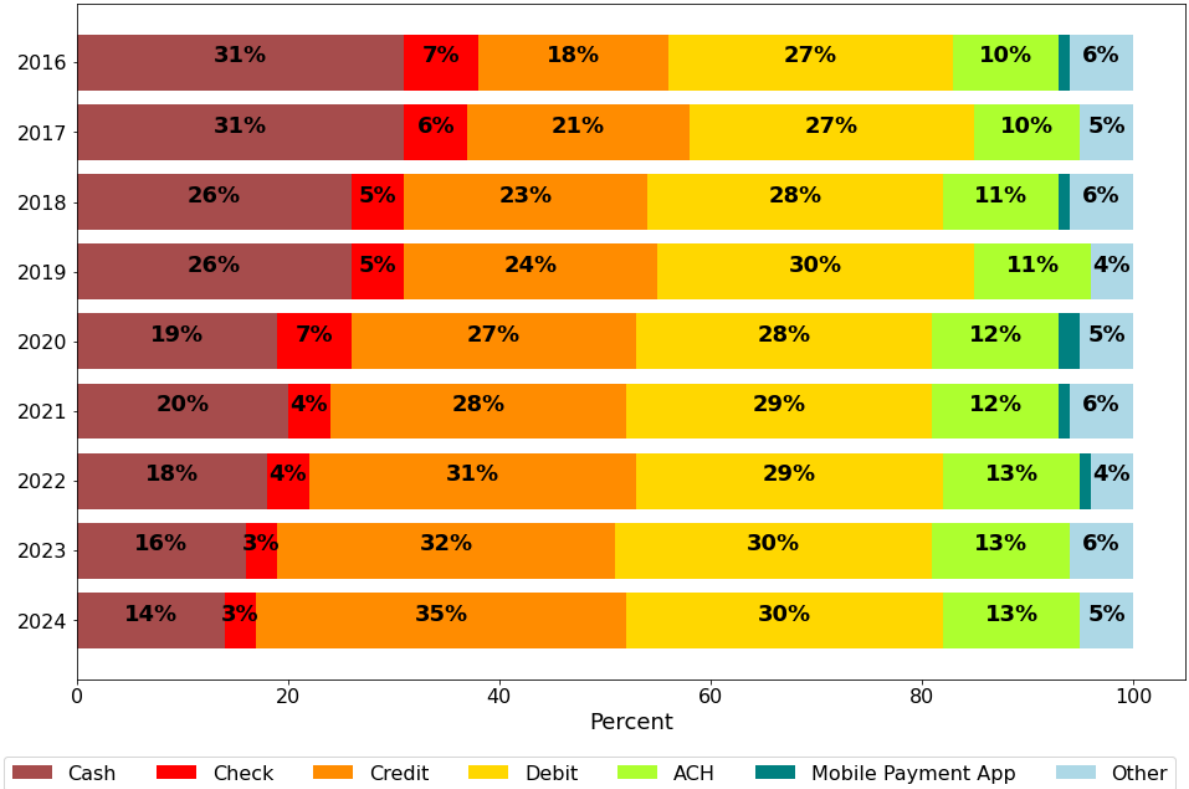

FIGURE 3   Share of different payment instruments used in all payments in the US.

*Notes*: ACH payments include bank account number payments, online banking bill pay, and some payments made through mobile payment apps. The source of the data presented in this figure is the 2025 Findings from the Diary of Consumer Payment Choice, the Federal Reserve Financial Services (Figure 2, page 5), https://www.frbservices.org/binaries/content/assets/crsocms/news/research/2025-diary-of-consumer-payment-choice.pdf (accessed September 18, 2025).

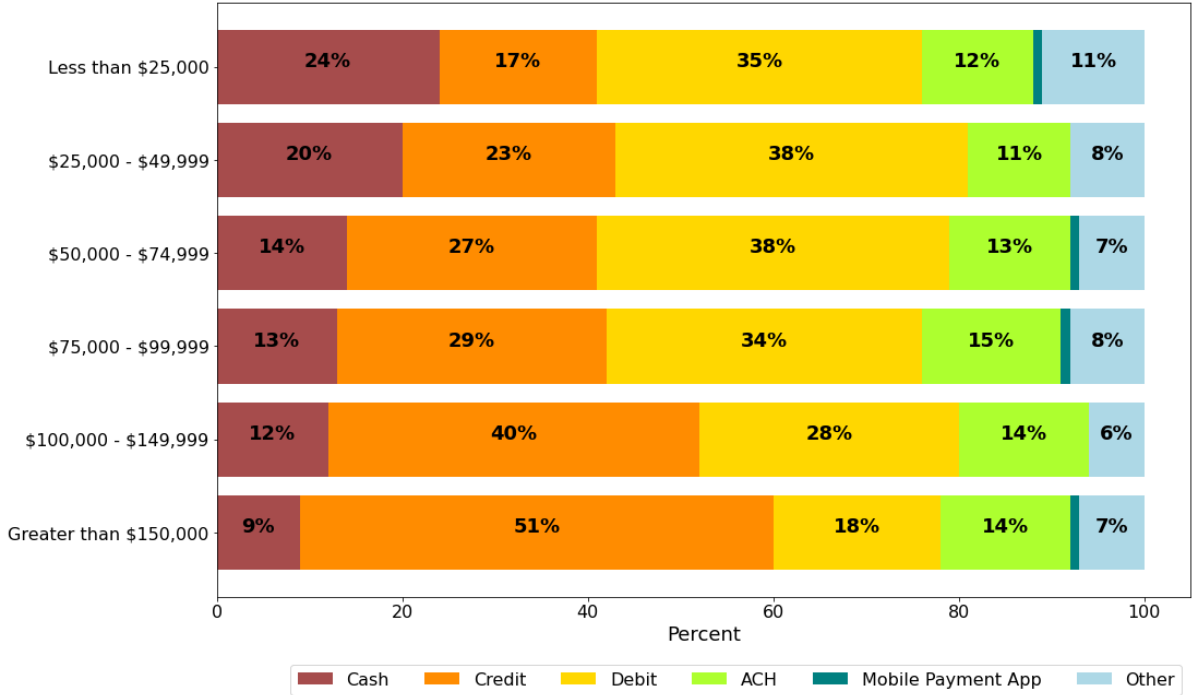

FIGURE 4    Share of different payment instruments' use by household income in the US.

*Note*: The source of the data presented in this figure is the 2025 Findings from the Diary of Consumer Payment Choice, The Federal Reserve Financial Services (Figure 10, page 13), https://www.frbservices.org/binaries/content/assets/crsocms/news/research/2025-diary-of-consumer-payment-choice.pdf (accessed September 18, 2025).

TABLE 1   Estimates of the rounding tax in 2013.

| | (1)<br>Average rounding tax per transaction (in NIS) | (2)<br>Revenue (%) | (3)<br>Transactions (in 1,000s) | (4)<br>Rounding tax (in NIS): Equal shares | (5)<br>Maximum rounding tax (in NIS) | (6)<br>Minimum rounding tax (in NIS) |
|---|---|---|---|---|---|---|
| Supermarkets and drugstores | 0.0075 | 83.80 | 188,856 | 353,962 | 150,371 | 422,390 |
| Small grocery stores | 0.0058 | 15.30 | 98,822 | 143,836 | 575,343 | 0.0 |
| Convenience stores | 0.0048 | 0.80 | 7,856 | 9,482 | 37,926 | 0.0 |
| **Total** | | **100.00** | **295,533** | **507,280** | **763,641** | **422,390** |

*Notes*: Column 1 gives the average amount paid by consumers per transaction. Column 2 gives the share of each store type in total FMCG revenue. Column 3 gives the annual number of FMCG transactions per store type. Column 4 gives the total annual rounding tax under the assumption that in all stores, cash transactions comprise 25% of all transactions. Column 5 (6) gives the maximum (minimum) possible rounding tax when we keep the total share of cash transactions at 25% but allow it to vary across store types. The maximum (minimum) rounding tax is obtained when the share of cash transactions in superstores and drugstores is 10.6% (29.8%), and 100% (0%) in small grocery stores and in convenience stores. Thus, the maximum (minimum) rounding tax is obtained when the rounding tax in supermarkets and drugstores (small grocery and convenience stores) is lower than in the baseline (equal shares) scenario.